# Analytical impact excitation of Er/O/B co-doped Si light emitting diodes

Xiaoming Wang[1], Jiajing He[1], Ao Wang[2], Kun Zhang[3], Yufei Sheng[2], Weida Hu[3], Chaoyuan Jin[4], Hua Bao[2], Yaping Dan[1*]

[1]University of Michigan – Shanghai Jiao Tong University Joint Institute, Shanghai Jiao Tong University, Shanghai 200240, China

[2]Global Institute of Future Technology, Shanghai Jiao Tong University, Shanghai 200240, China

[3]State Key Laboratory of Infrared Physics, Shanghai Institute of Technical Physics, Chinese Academy of Sciences 500 Yu Tian Road, Shanghai 200083, China

[4]College of Information Science and Electronic Engineering, Zhejiang University, Hangzhou 310007, China

Abstract

Er doped Si light emitting diodes may find important applications in the generation and storage of quantum information. These diodes exhibit an emission efficiency two orders of magnitude higher at reverse bias than forward bias due to impact excitation. However, physics of impact excitation in these devices remains largely unexplored. In this work, we fabricated an Er/O/B co-doped Si light emitting diode which exhibits a strong electro-luminescence by the impact excitation of electrons inelastically colliding the Er ions. An analytical impact excitation theory was established to predict the electroluminescence intensity and internal quantum efficiency which fit well with the experimental data. From the fittings, we find that the excitable Er ions reach a record concentration of $1.9 \times 10^{19}\,\mathrm{cm}^{-3}$ and up to 45% of them are in excitation state by impact excitation. This work has important implications for developing efficient classical and quantum light sources based on rare earth elements.

An integrated chip system that possesses the generation,[1] processing,[2-3] transmission and storage of quantum information[4] is vital for quantum computing, in particular if such a system can be integrable with the existing Si based complementary-metal-oxide-semiconductor (CMOS) circuitry. Er ions implanted in silicon provide an unprecedented opportunity to develop such a system due to several important features of Er ions.[5-7] Firstly, Er ions absorb or emit photons at communication wavelength that is compatible with the existing optical communication systems and the emerging silicon photonic technology.[8-9] Secondly, rare earth elements in crystals often have a long coherent time,[10] a striking feature that is crucial for quantum information processing. Er as one of the most extensively investigated rare earth elements is not an exception. Lastly, Er ions have a fine spin energy structure which can be coupled with electromagnetic wave in microwave or radio frequent range. As a result, the electron spin coherence in Er ions can be manipulated optically and electronically.[11-13]

In the past several decades, Er in various substances has been well studied including crystalline and nanostructured semiconductors and dielectrics.[14-18] The research on Er in oxide or other dielectrics is fruitful, partly because Er has a high solubility in these materials.[15-16] A high concentration of these Er ions can be directly excited by photons with the right energy. Optically pumped lasers and optical amplifiers based on Er doped fibers have been commercialized. The pursue of electrically pumped Er doped Si light emitting devices turns out to be futile due to the fact that Er ions in these devices are excited by the recombination of electrons and holes from the energy bands via Er-related defects in Si bandgap.[5] The concentration and energy level of these defects are difficult to engineer to maximize the Er excitation efficiency while maintaining a high concentration of Er ions optically active. As a result, the quantized energies transferred by the electron-hole recombination via these defects are often not in a good match for Er excitation, as a result of which emission efficiencies are extremely low in these devices.[10]

Impact excitation provides a new avenue to create efficient Er-doped Si light emitting diodes (LEDs).[19-20] Like optical pumping, impact excitation can provide tunable quantized energies by the electric field in the depletion region of the PN junction diode in which electrons are ballistically accelerated within the mean free path and inelastically collide to excite Er ions. Er doped Si LEDs under reverse bias were previously reported to have much higher luminescence intensities than forward bias,[21-

[23] which was believed to be caused by impact excitation. However, the physics of impact excitation has remained largely unexplored.

In this work, we fabricated an Er/O/B codoped Si LED that exhibits an electroluminescence (EL) emission two orders of magnitude stronger at reverse bias than forward bias. Near infrared camera and scanning photocurrent microscopy were employed to verify that the EL emission comes the impact excitation in the depletion region under a reverse bias greater than 3V. The derived impact excitation theory predicts that the emission rate and internal quantum efficiency follow the nonlinear functions of reverse bias, which, excitingly, was consistent with experimental observations. The fitting of the theory with experimental data allows us to find that the optical excitable Er concentration reaches a record of $1.8 \times 10^{19}$ cm$^{-3}$, 45% of which are excited at a maximum reverse bias of 11V.

**Results and Discussion**

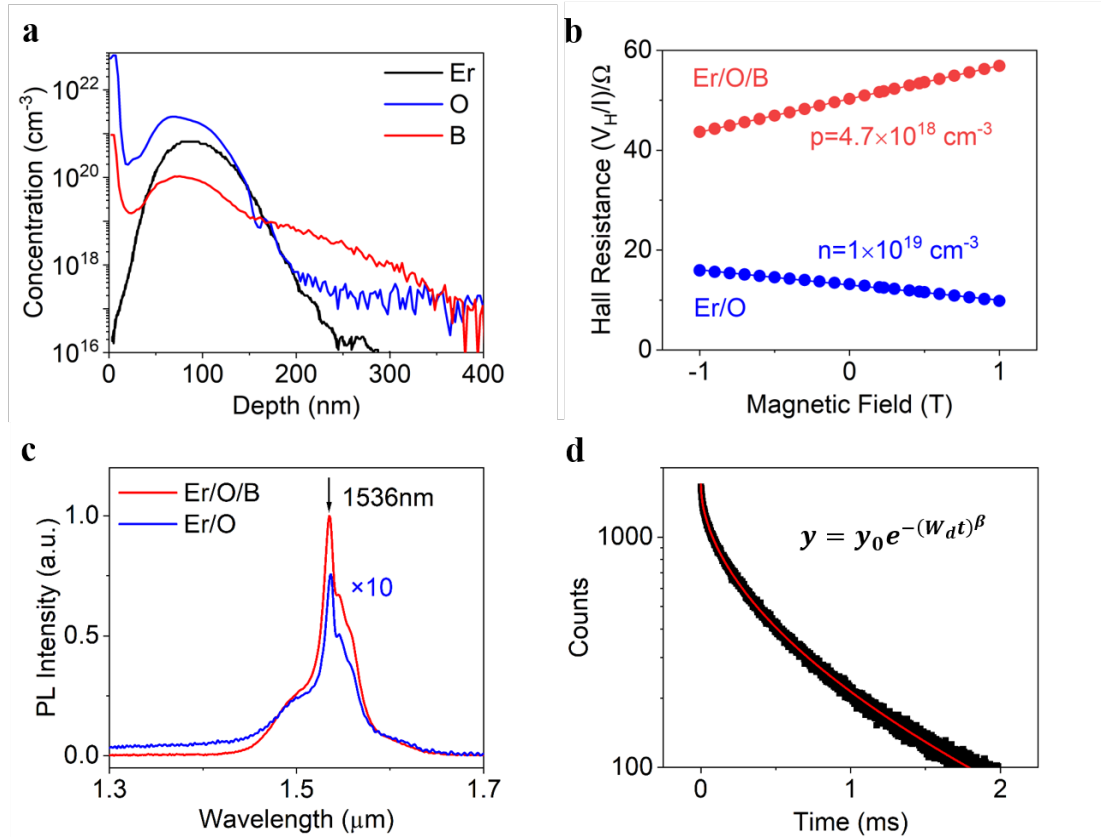

**Figure 1.** (a) SIMS results of Er (black), O (blue) and B (red) distribution. (b) Hall resistance dependence on the magnetic field at 300 K for ErO: Si and ErOB: Si samples. (c) PL spectra of ErO

doped Si and ErOB doped Si samples. **(d)** Transient PL decay which can be fitted with a stretched exponential function.

The intrinsic silicon wafer was first doped with Er and O and then treated with the deep cooling process in which the sample was rapidly quenched from high temperature (950°C) by flushing with Helium gas cooled with liquid nitrogen (77K). We previously demonstrated that the deep cooling process can efficiently suppress the precipitation of Er and reduce the concentration of nonradiative recombination centers,[24] as a result of which the Er/O doped Si has a two-order-of-magnitude enhancement of light emission efficiency. Properly tuning the concentration of Er and O (black and blue line in Fig.1a will further optimize the photoluminescence (PL) efficiency. The resultant Er/O co-doped Si sample is n-type and degenerate with an electron concentration of $1\times10^{19}$cm$^{-3}$ (Fig.1b). To create a PN junction light emitting diode, we turned part of the Er/O doped Si into p-type by implanting boron ions for compensate doping (Fig.1a). After doping, the sample went through the same process for the Er/O doped Si. The co-doping of B in Er/O Si will enhance the photoluminescence efficiency of Er/O doped Si. The resultant Er/O/B co-doped Si sample is p-type and degenerate with a hole concentration of $4.7 \times 10^{18}$ cm$^{-3}$ (Fig.1b). The PL spectrum of Er/O and Er/O/B doped Si are shown in Fig.1c. The PL intensity is enhanced by more than one order of magnitude after introducing Boron doping. Fig.1d plots the transient PL decay of the Er/O/B co-doped silicon sample. As we showed previously,[25] defect radiative decay transients in semiconductors can be best described by a stretched exponential decay function (Kohlrausch's function) $\emptyset(t) = \emptyset_0 \exp[-(W_d t)^\beta]$ in which $\emptyset_0$ is the steady-state PL photo flux density and $W_d$ is the effective radiative probability of Er ions. We used this stretched exponential decay function to fit the experimental decay transients and found $W_d = 4.3\times10^3$ s$^{-1}$.

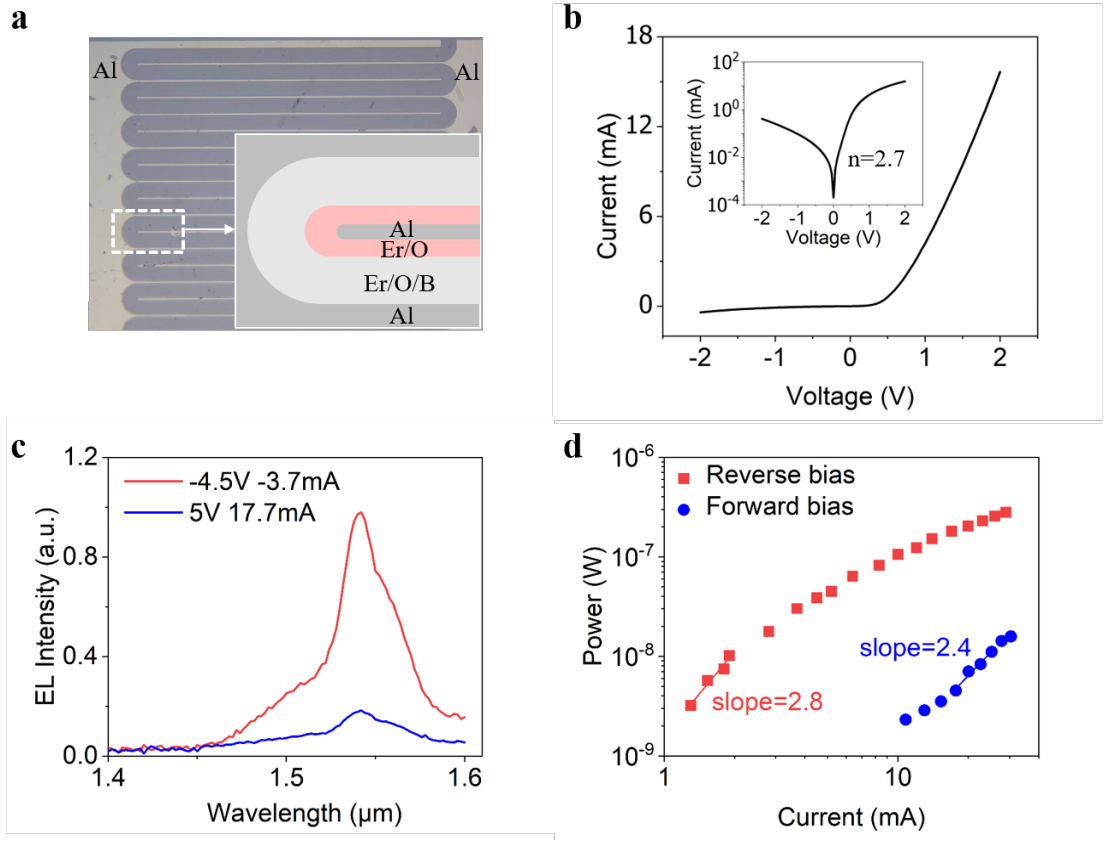

**Figure 2. (a)** Optical microscopic image of the device. Inset: A closeup sketch of PN junction. **(b)** Current vs Voltage (*I-V*) characteristics of the Er/O/B co-doped Si PN junction. Inset: *I-V* characteristics in semilogarithmic scale. **(c)** EL spectra under reverse and forward bias. **(d)** Photo emission power under the injection of different currents at room temperature.

To make electrically pumped light emitting diodes, we made a cross-finger shaped PN junction diode by introducing B dopants at the optimized concentration into the Er/O doped silicon as shown in Fig.2a. The dopants were activated by deep cooling process. A pair of cross-finger electrodes were deposited to contact the $n^+$-type Er/O and $p^+$-type Er/O/B region, respectively. The contacts are Ohmic and the contact resistances are negligible in comparison with the resistance of the PN junction at reverse bias (see SI section 1). The current vs voltage (*I-V*) characteristics of the PN junction diode are shown in Fig.2b. The leakage current is relatively large with only an order of magnitude lower than the forward current. The relatively large leakage current is likely due to the high concentration of defects by Er/O/B doping that significantly increases the thermal generation of electrons and holes in the depletion region. A higher reverse bias will further increase the leakage current, which will be investigated later. The ideality factor is found as ~ 2.7 from the slope of the logarithmic forward current at small forward bias (inset of Fig.2b). According to the device physics of PN junction diodes,[26] the forward current dominated by defect-mediated recombination in depletion region will lead to an ideality of ~ 2. The extracted ideality of ~ 2.7 is largely consistent with the fact that the high concentration of defects by Er/O/B doping in the depletion region allows for efficient recombination of electrons and holes injected under small forward bias.

The electroluminescence (EL) spectrum of the PN junction diode under forward (black dots) and reverse bias (red squares) is shown in Fig.2c. Both spectra exhibit a peak near 1.536 μm which is nearly identical to the recorded photoluminescence (PL) of Er/O co-doped Si in Fig.1c, confirming that the EL emission comes from Er ions. Fig.2d shows the integrated EL intensity under forward (black dots) and reverse current (red squares). The EL under reverse bias is nearly two orders of magnitude stronger than the forward bias. Interestingly, the dependence of EL intensity on pump current has a slope of 2.4-2.8, indicating that there is some stimulated emission inside the diode. We previously observed similar phenomenon in other Er doped Si light emitting diodes.[7]

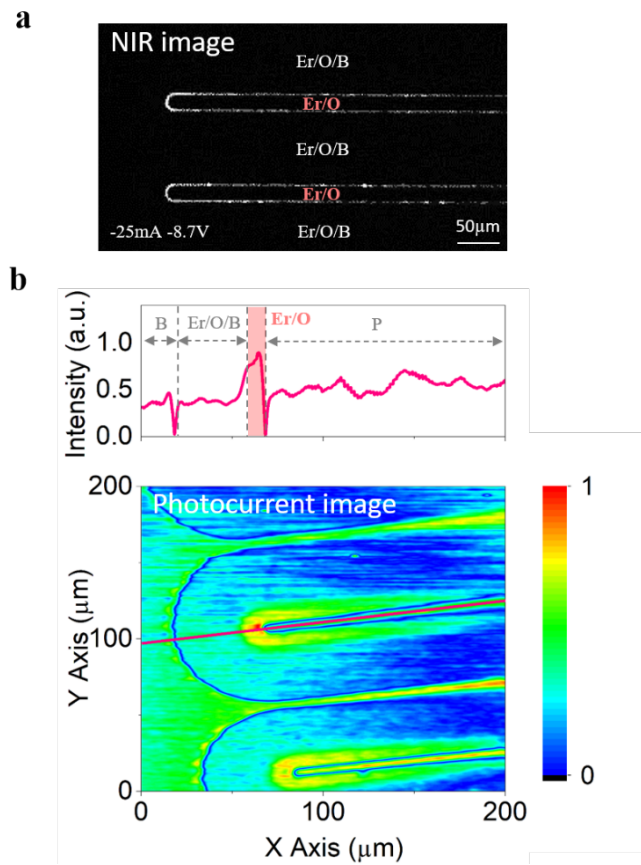

**Figure 3. (a)** Optical microscopic image in near infrared spectrum of the cross-finger PN junction

diode under reverse bias. **(b)** Scanning photocurrent microscopic image of the diode at 0V. Top: photocurrent profile along the red line. Bottom: photocurrent map. Normalized photocurrent is coded in color.

To investigate the exact location where the EL emission comes from, we employed a near-infrared (NIR) optical microscope (Fig.3a) to image the spatial distribution of EL emission. By comparing the visible optical image and NIR image, we identify that the EL emission comes from the intersection of the Er/O and Er/O/B region. The Hall effect measurements in Fig.1b have shown that the Er/O region is $n^+$-type and the Er/O/B region is $p^+$-type. This means that the EL emission likely comes from the depletion region of the PN junction diode. To verify this observation, we utilized a scanning photocurrent microscopy to visualize the built-in electric field in the diode. A photocurrent map is created by scanning a micrometer sized light spot (λ=532nm) focused by an optical microscope over the diode surface. When the light spot comes cross the built-in depletion region, the photogenerated electron-hole pairs will be efficiently separated and collected as photocurrent.[27] The photocurrent map in Fig.3b indicates that the photocurrent pattern in the map is nearly identical to the EL pattern in the NIR image (Fig.3a), except that some large photocurrent shows up near the metal electrode edge probably due to light scattering by the metal electrode.

Clearly, the reverse biased emission comes from the depletion region in which electrons are accelerated to inelastically collide the Er ions. The kinetic energy of electrons lost in the collision has a probability to transfer to the Er ions such that electrons in the 4f shell are excited from the ground state $^4I_{15/2}$ to the excited state $^4I_{13/2}$. The relaxation of the excited electrons will emit photons at λ=1.536 μm ($\hbar\omega = 0.8eV$) with a probability of $W_d$, generating a transient decay similar to Fig.1d.

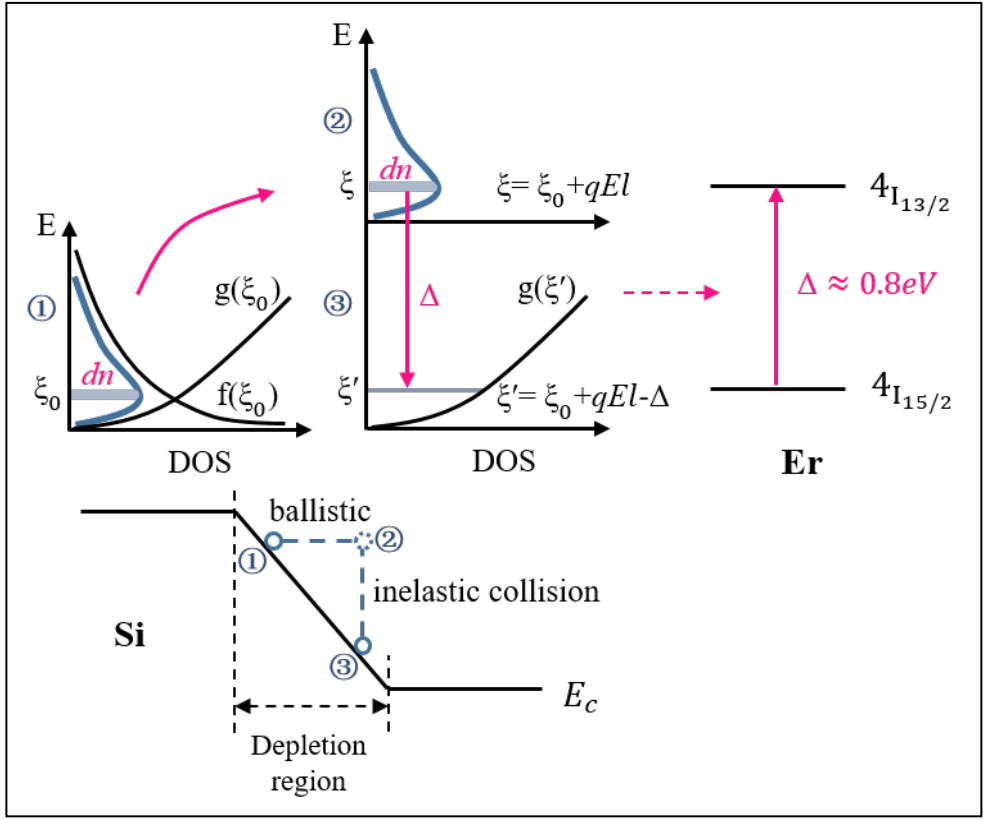

**Figure 4.** Impact excitation mechanism. Electrons in the location ① will be ballistically accelerated to the location ② in a short time (~tens of fs) such that the hot electrons in the location ② distribute identically to the location ①. After the inelastic collision, those electrons lose an energy of Δ that will be absorbed by Er ions to excite electrons from the ground state to excited state.

To theoretically investigate this energy transfer process, let us focus on electrons in the depletion region starting from the location ① in thermal equilibrium (see Fig.4). Within a distance of the mean free path, the electrons will ballistically transport until they collide the impurity ions (mostly Er ions). In our case, the depletion region width is estimated to be ~ 45 nm due to the lateral randomness of implanted ions (see SI section 2). EL from our PN junction diode was detectable when the reverse bias is greater than 3V. The maximum electric field intensity inside the depletion region is on an order of $10^6$ V/cm, meaning that some electrons in the depletion region have already reached the velocity saturation according to the electric field dependent velocity of electrons in silicon. Therefore, electrons will mostly inelastically collide with impurities at the location ② after travelling a distance of mean free path (~4 nm from our later results). Given the time of ballistic transport is on order of ~ tens of fs, electrons before collision are hot electrons with a distribution similar to ① but at a higher energy of $qEl$ with $q$ being the unit charge, $E$ the electric field intensity and $l$ the mean free path. After collision, electrons will relax to empty states at ③.

Electrons of Er ions at $^4I_{15/2}$ states are excited to $^4I_{13/2}$ states only if hot electrons at $\xi = \xi_0 + qEl$ in ② are relaxed by inelastic collision to states at $\xi'$ (mostly empty in ③) by losing an energy of $\Delta$ ($\cong 0.8$ eV). For all electrons in the location ②, the probability $W_i$ of an Er ion being excited is given by the equation (1) according to previous theoretical work.[28]

$$W_i = W_0 \iint_{\xi_0 \geq 0, \xi' \geq 0} g(\xi) f(\xi) g(\xi') \delta(\xi - \Delta - \xi') d\xi d\xi' \quad (1)$$

, in which $\xi = \xi_0 + qEl$, $W_0$ is associated with the Er ion radius and Coulombic interactions ($1.4 \times 10^{-9}$ cm$^{-3}$ eV$^{-1/2}$ s$^{-1}$),[28] $g$ is the density of states and $f$ is the carrier distribution function. Since the free carrier concentration in the depletion region is relatively low, $f$ will follow the Boltzmann distribution although the Si is degenerate due to high doping concentrations. The last term $\delta(\xi - \Delta - \xi')$ in the integral is to ensure that the energy loss in the inelastic collision matches the energy needed for electrons in Er ion to excite from $^4I_{15/2}$ to $^4I_{13/2}$ states.

Depending on whether the kinetic energy is higher or lower than the energy difference $\Delta$, the probability of impact excitation $W_i$ given in eq.(2) can be derived from eq.(1) (see SI section 3)[28] where $n$ is the hot electron concentration, $k$ is the Boltzmann constant and $T$ is the electron temperature. In our case, the kinetic energy of electrons is higher than 0.8 eV (see later results). Therefore, we only focus the case $qEl \geq \Delta$ in eq.(2) for the following derivation.

$$W_i = \begin{cases} \frac{W_0}{\pi} n \sqrt{qEl - \Delta} & qEl \geq \Delta \\ \frac{W_0}{\pi} n \sqrt{\Delta - qEl}\, e^{-\frac{\Delta - qEl}{kT}} & qEl \leq \Delta \end{cases} \quad (2)$$

Note that the electric field intensity and dopants in our PN junction is not spatially uniform, and that the emission probability $W_i$ is dominated by the maximum electric field intensity. For simplicity, we choose to use the maximum electric field intensity $E_m$ in eq.(2) for a given reverse bias $V_a$, which is written as $E_m = \frac{2(V_{bi} + V_a)}{W_{dep}}$ with $V_{bi}$ being the built-in potential (~ 1.06 V), and $W_{dep}$ is the depletion region width. $W_{dep}$ is ~ 45 nm estimated from the lateral randomness of ion implanted dopants and regarded as constant due to the fact that both Er/O/B and Er/O doped Si are degenerate (see SI section 2).

In eq.(2), the electron concentration $n$ is also dependent on the reverse bias $V_a$. Since the electrons in depletion region under a reverse bias of 3V have already reached the velocity saturation in our device, the electron concentration $n$ can be expressed in terms of leakage current density $J$ and electron saturation velocity $v_s$ as $n = \frac{J}{qv_s}$. The experimental leakage current and then the electron concentration can be fitted with an exponential (or polynomial) function of reverse bias $V_a$ as shown in Fig. 5a. After plugging the analytical dependence of $n$ and $E$ on $V_a$ into eq.(2), we manage to find the analytical expression of $W_i$ dependent on $V_a$ which is plotted in Fig.5b.

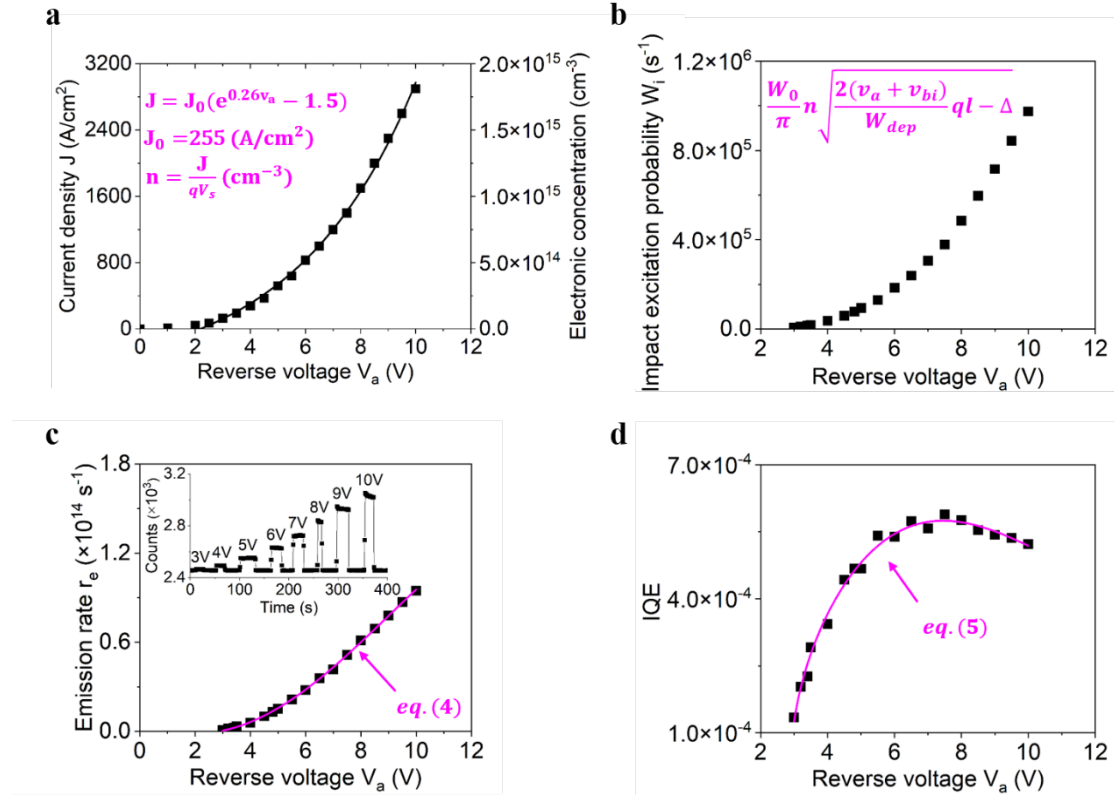

**Figure 5. (a)** Experimental leakage current density that is exponentially dependent on reverse bias $V_a$ which can be fitted with an empirical function. **(b)** Photo emission probability due to impact excitation by plugging the measured electron concentration in (a) to eq.(2). **(c)** Emission rate as a function of reverse bias $V_a$. Inset: EL under a pulsed bias to mitigate thermal effect. **(d)** Experimentally measured *IQE* that is well fitted with the theoretical *IQE* in eq.(5).

Note that the kinetic energy of inelastic collision can be only transferred to Er ions that are not in excitation state (in ground state). Let us suppose the concentration of all optically excitable Er ions and those already in excitation state are $N_{tt}$ and $N_{Er}$, respectively. Er ions in ground state will be $N_{tt} - N_{Er}$ in concentration. The radiative emission probability of Er ions in excitation state is $W_d$. At steady state, the excitation rate of Er ions in ground state by inelastic collision of hot electrons will be equal to the relaxation rate of Er ions from excitation state to ground state, which can be described by eq.(3).

$$W_i(N_{tt} - N_{Er}) = N_{Er}W_d \tag{3}$$

From eq.(3), we find that the concentration of Er ions in excitation state $N_{Er}$ with the expression of $W_i$ given in eq.(2). The photon emission rate $r_e$ from our Er doped Si is given by $r_e = N_{Er}W_d L_{ex} A_c$ where $A_c$ is the cross-sectional area of the conduction channel with the cross finger interface length of 1cm and depth of 100 nm according to the doping profile in Fig.1a. $L_{ex}$ is the length of the optically active region (should be around where the maximum electric field intensity is). The full expression of $r_e$ is given as a function of reverse bias $V_a$ in eq.(4) after plugging $N_{Er}$ found from eq.(3), the empirical expression of the electron concentration in Fig.5a and $W_i$ given in eq.(2) along with the maximum electric field intensity.

$$r_e = \frac{W_i}{W_i + W_d} N_{tt} W_d L_{ex} A_c = \frac{J\sqrt{V_a+V_{bi}-A}}{J\sqrt{V_a+V_{bi}-A} \ + \ B/J_0} N_{tt} W_d L_{ex} A_c \tag{4}$$

, in which $A = \frac{W_{dep}}{2ql}\Delta$ and $B = \frac{\pi v_s W_d}{W_0}\sqrt{\frac{qW_{dep}}{2l}}$.

To quantitatively measure the photon emission rate $r_e$ from our reversely biased PN junction diode, we used a commercially calibrated near-infrared light emitting diode (InGaAs diode L12509-0155G) as a reference to quantify the light collection efficiency of our optical system (see SI section 4 for details). Silicon has a relatively high refractive index. Light emitted from Si internally may be trapped inside. Based on finite difference time domain (FDTD) simulations, we found the light extraction efficiency from silicon to air is estimated as 2.3% (see SI section 5). With these considerations, we recorded the photon emission rate $r_e$ from our device at pulsed bias to mitigate the thermal effect, as shown in Fig.5c along with the inset.

The experimentally measured photon emission rate can be fitted with the theoretical rate given in eq.(4). From the fitting, we extracted three important correlations given $V_{bi}$ = 1.06 V (SI section 2). Firstly, $A = \frac{W_{dep}}{2ql}\Delta$ = 4.3 with $l$ being the mean free path and $\Delta$ = 0.8 eV, from which we find $\frac{W_{dep}}{l}$ = 11. As $W_{dep}$ can be reliably estimated as ~ 45 nm from the SRIM simulation of dopant distribution (SI section 2), we conclude that the electron mean free path in our sample is ~ 4 nm, which is consistent with the results in literature.[29] Secondly, $\frac{B}{J_0} = \frac{\pi v_s W_d}{W_0 J_0}\sqrt{\frac{qW_{dep}}{2l}}$ = 31 $V^{-1/2}$ from which we find $W_d = 1 \times 10^6 \ s^{-1}$ given $J_0$ = 255 A/cm² from the experimental current fitting in Fig.5a, $W_0 = 1.4 \times 10^{-9}$ cm$^{-3}$ eV$^{-1/2}$ s$^{-1}$ given in eq.(1) and $v_s$ the electron saturation velocity often equal to ~ $10^7$ cm/s in silicon.[30] The extracted $W_d$ is 2-3 orders of magnitude larger than what we found optically in Fig.1d. This is either because strong electric fields can enhance the emission probability of excited Er ions or due to the overestimation of leakage current. The third important correlation is $N_{tt}W_d L_{ex} A_c = 1.9 \times 10^{14}$ s$^{-1}$ in which $L_{ex}$ is the length of optically active region. The depletion region is ~ 40 nm wide and $L_{ex}$ is part of it. On the other hand, $L_{ex}$ should be longer than the electron mean-free path which is ~ 4 nm. Therefore, it is reasonable to set $L_{ex}$ as ~ 10 nm. Based on these estimated parameters, we find $N_{tt} = 1.9 \times 10^{19}$ cm$^{-3}$. Fig.1a shows the maximum Er doping concentration is ~ $7 \times 10^{20}$ cm$^{-3}$. This indicates that only 2.5% of implanted Er ions are optically active, which is consistent with our previous finding by optical excitation.[25] Among these $1.9 \times 10^{19}$ cm$^{-3}$ optically excitable Er ions, up to $8.1 \times 10^{18}$ cm$^{-3}$ (~ 45 %) are in excited state at a maximum reverse bias of $V_a = 11$V. To summarize, we listed in Table I some known parameters from other sources and those extracted from our fittings.

Table I Parameters extracted by fitting the impact excitation theory with experimental results.

| Found by | Parameters | Physical meaning | value |
| --- | --- | --- | --- |
| Reference | $v_s$ | Electron saturation velocity | ~$10^7$ cm/s |
| Estimation | $L_{ex}$ | Length of optically active region | ~ 10 nm |
| Simulations | $W_{dep}$ | Depletion region width | ~ 45 nm |
| Measurements | $A_c$ | Cross-sectional area | ~ $10^{-5}$ cm$^{-2}$ |
| Fitting extraction | $N_{tt}$ | Optically excitable Er ions | $1.8 \times 10^{19}$ cm$^{-3}$ |
| | $W_d$ | Emission probability of excited Er ions | $1 \times 10^6$ s$^{-1}$ |
| | $N_{Er}$ (11V) | Er ions in excitation state at reverse bias of 11V | $8.1 \times 10^{18}$ cm$^{-3}$ |
| | $l$ | Mean free path (nm) | ~ 4 nm |
| | $\eta$ | Ratio of optical active Er ions to all Er ions | ~ 2.5% |

The number of electrons injected into the reverse biased PN junction diode per second is given by $JA_c/q$ where $J$ is the leakage current density, $A_c$ is the cross-section area of the conduction channel and $q$ is the unit of charge. The internal quantum efficiency IQE is given by the ratio of $r_e$ to $JA_c/q$ as shown in eq.(5).

$$IQE = \frac{N_{Er}W_dL_{ex}A_c}{JA_c/q} = \frac{\frac{W_0}{\pi}n\sqrt{qE_ml-\Delta}}{\frac{W_0}{\pi}n\sqrt{qE_ml-\Delta}+W_d} \frac{qN_{tt}W_dL_{ex}}{J} \quad (5)$$

, in which $L_{ex}$ is the length of optically active region and $v_s$ is the saturation velocity of electrons. IQE in eq.(5) becomes an explicit function of $V_a$ after plugging the empirical expression of $J$ given in Fig.5a with $n = \frac{J}{qv_s}$ and the maximum electric field intensity dependent on $V_a$.

Fig.5d plots the experimentally measured IQE as a function of reverse bias voltage. IQE increases from 3 V and starts to drop after reaching the maximum at ~ 7 V. Amazingly, this nonlinear dependence of IQE on $V_a$ can be nicely fitted with eq.(5). From the fitting, we extracted the same parameters with the previous fitting of the theoretical emission rate with the experimental rate in Fig.5c.

In the end, we need to point out that lateral surface devices often have significant surface leakage current, in particular at a large reverse bias in our case. It means that the leakage current passing through the depletion region of our device is likely overestimated. This overestimation does not affect the measurement of EL intensity but underestimate the IQE presented in Fig.5d, resulting in the $N_{tt}$ underestimated and $N_d$ overestimated. The fabrication of vertical PN junction diodes will effectively mitigate the surface leakage current and increase the IQE. In addition to suppressing the surface leakage current, increase of optically active Er ions in the depletion region also helps to enhance the IQE. One possible approach is to reduce the free carrier concentration of Er/O/B doped Si by fine-tuning the Boron dose so that this Er/O/B region can has a low concentration of free charge carriers. Under reverse bias, the whole quasi-intrinsic Er/O/B region can be depleted with a high electric field inside, which will significantly increase the emission efficiency. On the other hand, we can reduce the concentration of Er ions in the optically active region. A single Er ion may be potentially excited to emit single photons each time. Such a single photon source can be monolithically integrated with silicon waveguide, modulators, photodetectors and CMOS transistors for developing silicon quantum photonics.

**Conclusion**

In this work, we fabricated an Er doped Si light emitting diode with Er/O and Er/O/B as n$^+$ and p$^+$

region, respectively. The EL emission at 1.536 μm from the depletion region of the Si diode is enhanced by two orders of magnitude under reverse bias. The internal quantum efficiency (IQE) exhibits a nonlinear dependence on reverse bias, which increases with reverse bias but starts to decline after reaching a maximum IQE of 0.06% at 7V. To theoretically analyze the EL emission, we developed an impact excitation theory for hot electrons with high kinetic energy to excite Er ions by inelastic collision. This theory allows us to establish a theoretical IQE that fits well with the experimentally measured nonlinear IQE. From the fitting, we find that 2.5% implanted Er ions are optically active with a record concentration reaching $1.9 \times 10^{19}$ cm$^{-3}$. Nearly 45% of the optically active Er ions are excited by impact excitation at a reverse of 11V. The low IQE is likely because the leakage current involved in the impact ionization in our diode is overestimated due to the surface leakage current. Although our theory is derived for Er ions in Si, it can be generalized for emission from rare earth elements in semiconductors by impact excitation.

**Experimental**

**Ion implantation**

Float zone (FZ) intrinsic Si (100) wafers (resistivity ≥ 10 kΩ cm; thickness: 525 ± 10 μm; Sibranch, China) were first implanted with Er and O ions at the Institute of Semiconductors, Chinese Academy of Sciences in Beijing, China. The injection energies and dosages of Er and O ions were 200 keV, $4 \times 10^{15}$ cm$^{-2}$, and 30 keV, $1 \times 10^{16}$ cm$^{-2}$, respectively. Boron was implanted in the pre-defined region at Shanghai Institute of Technical Physics, Chinese Academy of Sciences. The injection energy and dosage of B ions was 20 keV and $1 \times 10^{15}$ cm$^{-2}$, respectively.

**Device fabrication**

Deep cooling process was applied to activate the implanted atoms and repair the lattice damage caused by ion implantation. The sample was first annealed 950 °C for 25 mins followed by a flush of high purity Helium (99.999 %) gas cooled by liquid nitrogen (77K), resulting in the temperature decreasing to -125°C in 5 s. In the end, the temperature of the sample slowly arose to room temperature.

Photolithography was applied to define the pattern of metal electrodes (AEMD, Shanghai Jiao Tong University). The residual photoresist was removed with oxygen plasma. The sample was then dipped in HF to remove SiO$_x$ and covered with 200nm thick Al film by thermal evaporation. After the liftoff process, the metal electrodes were wire-bonded to the PCB board for test.

**Optical measurements**

The PL was excited by a 405 nm laser spot with a diameter of 2 mm. PL spectra were recorded using an Edinburgh FLS1000 Spectrometer with a nitrogen-cooled near-infrared InGaAs photomultiplier tube. An LED control circuit was built using a npn power transistor to generate pulsed current. EL spectra were recorded using the monochromator (1132-1013-IHR320, Horiba) and an electrically cooled InGaAs detector. NIR imaging was measured by an InGaAs short-wave infrared camera (IH320-15; spectral response range: 0.9-1.7 μm; pixel:320×256; pixel distance:15 μm).

**Acknowledgement**

This work was financially supported by the special-key project of Innovation Program of Shanghai Municipal Education Commission (No. 2019-07-00-02-E00075), National Science Foundation of China (NSFC, No. 92065103), Oceanic Interdisciplinary Program of Shanghai Jiao Tong University (SL2022ZD107), Shanghai Jiao Tong University Scientific and Technological Innovation Funds (2020QY05) and Shanghai Pujiang Program (22PJ1408200). The devices were fabricated at the center for Advanced Electronic Materials and Devices (AEMD), and XPS measurements were carried out at the Instrumental Analytical Center (IAC), Shanghai Jiao Tong University.

**References**

[1] Z. Zhou, B. Yin, J. Michel, *Light: Science & Applications* **2015**, 4, e358.
[2] C. D. Bruzewicz, J. Chiaverini, R. McConnell, J. M. Sage, *Applied Physics Reviews* **2019**, 6.
[3] J. Wang, F. Sciarrino, A. Laing, M. G. Thompson, *Nature Photonics* **2020**, 14, 273.
[4] G. Zhang, J. Y. Haw, H. Cai, F. Xu, S. Assad, J. F. Fitzsimons, X. Zhou, Y. Zhang, S. Yu, J. Wu, *Nature Photonics* **2019**, 13, 839.
[5] A. Kenyon, *Semiconductor Science and Technology* **2005**, 20, R65.
[6] Y. Liu, Z. Qiu, X. Ji, A. Lukashchuk, J. He, J. Riemensberger, M. Hafermann, R. N. Wang, J. Liu, C. Ronning, *Science* **2022**, 376, 1309.
[7] J. Hong, H. Wen, J. He, J. Liu, Y. Dan, J. W. Tomm, F. Yue, J. Chu, C. Duan, *Photonics Research* **2021**, 9, 714.
[8] B. Wang, P. Zhou, X. Wang, *Applied Sciences* **2022**, 12, 11712.
[9] J. C. Palais, *Englewood Cliffs* **1984**.
[10] X. Wang, J. He, S. Jin, H. Liu, H. Li, H. Wen, X. Zhao, R. Abedini-Nassab, G. Sha, F. Yue, *Advanced Photonics Research* **2022**, 3, 2200115.

[11] H.-J. Lim, S. Welinski, A. Ferrier, P. Goldner, J. Morton, *Physical Review B* **2018**, 97, 064409.

[12] A. Ortu, A. Tiranov, S. Welinski, F. Fröwis, N. Gisin, A. Ferrier, P. Goldner, M. Afzelius, *Nature materials* **2018**, 17, 671.

[13] M. Le Dantec, M. Rančić, S. Lin, E. Billaud, V. Ranjan, D. Flanigan, S. Bertaina, T. Chanelière, P. Goldner, A. Erb, *Science advances* **2021**, 7, eabj9786.

[14] A. Kenyon, *Progress in Quantum Electronics* **2002**, 26, 225.

[15] H. Sun, L. Yin, Z. Liu, Y. Zheng, F. Fan, S. Zhao, X. Feng, Y. Li, C.-Z. Ning, *Nature Photonics* **2017**, 11, 589.

[16] M. Miritello, R. Lo Savio, F. Iacona, G. Franzò, A. Irrera, A. M. Piro, C. Bongiorno, F. Priolo, *Advanced Materials* **2007**, 19, 1582.

[17] M. A. Lamrani, M. Addou, Z. Sofiani, B. Sahraoui, J. Ebothe, A. El Hichou, N. Fellahi, J. Bernede, R. Dounia, *Optics communications* **2007**, 277, 196.

[18] F. Priolo, G. Franzo, S. Coffa, A. Polman, S. Libertino, R. Barklie, D. Carey, *Journal of applied physics* **1995**, 78, 3874.

[19] G. Franzo, F. Priolo, S. Coffa, A. Polman, A. Carnera, *Applied physics letters* **1994**, 64, 2235.

[20] S. Harako, S. Yokoyama, K. Ide, X. Zhao, S. Komoro, *physica status solidi (a)* **2008**, 205, 19.

[21] G. Franzo, S. Coffa, F. Priolo, C. Spinella, *Journal of Applied Physics* **1997**, 81, 2784.

[22] G. Franzo, F. Priolo, S. Coffa, A. Polman, A. Carnera, *Nuclear Instruments and Methods in Physics Research Section B: Beam Interactions with Materials and Atoms* **1995**, 96, 374.

[23] B. Zheng, J. Michel, F. Ren, L. Kimerling, D. Jacobson, J. Poate, *Applied Physics Letters* **1994**, 64, 2842.

[24] H. Wen, J. He, J. Hong, S. Jin, Z. Xu, H. Zhu, J. Liu, G. Sha, F. Yue, Y. Dan, *Advanced Optical Materials* **2020**, 8, 2000720.

[25] H. Liu, U. Kentsch, F. Yue, A. Mesli, Y. Dan, *Journal of Materials Chemistry C* **2023**, 11, 2169.

[26] C.-T. Sah, R. N. Noyce, W. Shockley, *Proceedings of the IRE* **1957**, 45, 1228.

[27] Y. Dan, K. Seo, K. Takei, J. H. Meza, A. Javey, K. B. Crozier, *Nano letters* **2011**, 11, 2527.

[28] I. Yassievich, L. Kimerling, *Semiconductor science and technology* **1993**, 8, 718.

[29] B. Qiu, Z. Tian, A. Vallabhaneni, B. Liao, J. M. Mendoza, O. D. Restrepo, X. Ruan, G. Chen, *Europhysics Letters* **2015**, 109, 57006.

[30] K. Seeger, *Semiconductor physics*, Springer Science & Business Media, **2013**.